# Enhancing Optimal Microgrid Planning with Adaptive BESS Degradation Costs and PV Asset Management: An Iterative Post-Optimization Correction Framework

Hassan Zahid Butt, and Xingpeng Li

*University of Houston, 4222 Martin Luther King Blvd, Houston, 77204, Texas, USA*

**Abstract**

The transition to renewable energy has positioned photovoltaic (PV) systems and battery energy storage systems (BESS) as essential assets in microgrids, particularly for remote installations. However, traditional planning models often neglect dynamic degradation costs or rely on complex or non-linear approaches, limiting their scalability and practical applicability. This paper introduces a microgrid planning model that integrates adaptive degradation cost modeling to enable accurate, efficient, and scalable long-term resource allocation. The proposed model employs the iterative post-optimization correction (IPOC) framework, solving a sequence of mixed-integer linear programming problems. Each iteration refines BESS degradation costs based on observed depth-of-discharge profiles and incorporates PV degradation costs to ensure realistic asset performance assessments. Sensitivity analysis of PV and BESS capital costs further underscores the model's robustness under varying economic conditions, with the IPOC framework achieving up to ~1% additional cost savings for the given test system compared to static approaches. The results demonstrate that by iteratively adjusting degradation penalties based on actual usage, the methodology optimizes BESS performance, ensures precise resource allocation, resolves issues of under- or overutilization, enhances system reliability, and facilitates scalable, sustainable microgrid planning.

## Nomenclature

Sets:

| | |
|---|---|
| D | Set of representative days in a single year |
| $T$ | Set of hourly time periods in a single day |
| $C$ | Set of BESS's cycle lives at specific DODs |

Indices:

| | |
|---|---|
| $t$ | Time period t, an element of set T |
| $d$ | Day d, an element of set D |

Parameters:

| | |
|---|---|
| $P_{d,t}^{load}$ | Total load (MW) in day d and hour t |
| $P_{d,t}^{PV}$ | Solar power capacity factor in day d and hour t |
| $C_{DG}^{op}$ | DG operational cost factor ($/MW) |
| $C_{DG}^{nl}$ | DG no-load cost ($/h) |
| $C_{DG}^{capital}$ | DG capital cost factor ($/MW) |
| $C_{PV}^{capital}$ | PV capital cost factor ($/MW) |
| $C_{BESS}^{capital}$ | BESS capital cost factor ($/MWh) |
| $\delta_{BESS}^{deg}$ | BESS degradation cost factor ($/MWh) |
| $\gamma_{PV}^{rep}$ | PV replacement cost as a percentage of capital cost |
| $\gamma_{BESS}^{rep}$ | BESS replacement cost as a percentage of capital cost |
| $T_{BESS}^{chg}$ | Duration of BESS charging (h) |
| $T_{BESS}^{dchg}$ | Duration of BESS discharging (h) |
| $P_{DG}^{min}$ | DG minimum output power (MW) |
| $U_{DG}^{init}$ | Initial commitment status of DG |
| $\eta_{BESS}$ | BESS roundtrip cycle efficiency |
| $SOC_{max}$ | Maximum state of charge limit for BESS |
| $SOC_{min}$ | Minimum state of charge limit for BESS |
| $DOD$ | BESS depth of discharge |
| $CL_{cycles}^{BESS}$ | BESS cycle life at various levels of DOD |
| $\delta_{PV}^{deg}$ | PV degradation rate per annum |
| $BigM$ | A very large number |
| $Y_{MG}$ | Total microgrid planning years |
| $\alpha$ | Scaling factor denoting the repetition of load and solar profile. |
| $N_D$ | Total number of days in the simulation resolution |
| $N_T$ | Total number of time periods per day in the simulation resolution |

Variables:

| | |
|---|---|
| $P_{DG}^{max}$ | DG maximum output power (MW) |
| $S_{PV}$ | PV system size (MW) |
| $S_{BESS}$ | BESS energy capacity (MWh) |
| $P_{d,t}^{DG}$ | DG output power (MW) in day d and hour t |
| $P_{d,t}^{pv_{curt}}$ | Solar power curtailed (MW) in day d and hour t |
| $P_{d,t}^{chg}$ | BESS charge power (MW) in day d and hour t |
| $P_{d,t}^{dchg}$ | BESS discharge power (MW) in day d and hour t |
| $E_{BESS}^{init}$ | Initial BESS energy level (MWh) |
| $E_{d,t}^{BESS}$ | Energy level of BESS (MWh) in day d and hour t |
| $C_{PV}^{deg}$ | PV degradation cost ($) |
| $C_{BESS}^{deg}$ | BESS degradation cost ($) |
| $U_{d,t}^{chg}$ | BESS charging status in day d and hour t |
| $U_{d,t}^{dchg}$ | BESS discharging status in day d and hour t |
| $U_{d,t}^{DG}$ | DG commitment status in day d and hour t |

## 1. Introduction

The shift toward renewable energy sources (RES) has elevated the significance of microgrids (MGs) in modern energy systems [1]. MGs enhance the resilience and efficiency of power systems through decentralized energy generation, distribution, and consumption. They can operate independently or in conjunction with the main grid, ensuring energy reliability during natural disasters or grid failures, particularly for critical facilities such as hospitals, emergency services and essential infrastructure [2]. Additionally, MGs support RES integration, reducing dependence on centralized power plants, promoting sustainability, and extending energy access to remote or underserved areas.

Among RES options, solar photovoltaic (PV) systems have become the most widely adopted due to their reduced costs, decentralized capabilities, and minimal maintenance requirements [3]-[4]. However, the intermittent nature of solar energy necessitates energy storage solutions for effective integration, particularly in remote MG setups [5]. Battery energy storage systems (BESS) have emerged as a top candidate for this role, offering high energy density, scalability, and ease of deployment compared to alternatives [6]-[7]. In conjunction with PV systems, BESS enhances reliability, sustainability, and energy optimization in MG environments.

Despite these benefits, the high upfront costs of PV systems and BESS remain a barrier to investment in MG projects [8]. Additionally, the degradation of both PV systems and BESS over time reduces efficiency and adds financial concerns related to maintenance and replacement [9]-[11]. While various types of distributed generation (DG) sources can be employed in MGs for energy resiliency, particularly in off-grid MG setups where no external grid is available as a backup, natural gas (NG) generators are specifically highlighted in this study due to their low operational costs, fast ramping capabilities, negligible start-up and shut-down costs, and minimal up/down time requirements [12]. These attributes make NG generators a practical choice for hybrid MG configurations, especially when integrating variable renewable resources like PV. The aim of this study is to address critical gaps in microgrid planning by incorporating degradation cost modeling for PV and BESS systems, using the latest technical and financial data.

### *1.1 Literature Review*

Many studies have addressed the complex problem of optimal BESS sizing for MG applications. A novel method was introduced in [13] for BESS sizing in a MG setup to reduce the system operational cost. A multiple objective optimization model using genetic algorithm was proposed in [14] for PV, BESS, diesel generators, and wind turbine sizing in a grid-independent MG. Firefly optimization algorithm was proposed in [15] for the sizing of a BESS system. The degradation cost associated with the BESS, however, has not been considered. BESS optimal sizing problem was tackled using particle swarm optimization method in [16], in which demand response was considered to stabilize MG frequency. Optimal allocation and sizing of a PV-BESS system on an IEEE 69-bus system was studied in [17]. In [18], optimal BESS sizing was computed for power quality and resiliency in MG.

The use of BESS for behind-the-meter stackable service applications and optimal sizing is discussed in [19]. Notably, this research addressed various aspects but did not delve into the replacement cost due to capacity degradation. A novel approach for optimal BESS sitting in transmission networks to reduce daily energy generation costs was proposed in [20]. A MG planning model is presented in [21] for determining distributed energy generation resource type and optimal generation setpoints albeit without the consideration of BESS's technical characteristics. Optimal BESS sizing for network congestion relief applications was explored in [22].

Meta-heuristic optimization techniques such as genetic algorithms and gravitation search were employed in [23] to tackle the optimal BESS sizing problem. The study included a performance comparison between multiple techniques, as well as an MG cost comparison with and without the inclusion of BESS revealing a 70% cost saving with BESS. However, important battery characteristics for degradation modeling were not incorporated. BESS size optimization for grid-connected and islanded MGs based on the operation cost was investigated in [24], but the degradation was also ignored.

A new mathematical formulation proposed in [25] explored the BESS's role in a distribution network but overlooked the relationship between BESS cycle life and depth of discharge (DOD) in the computation of optimal BESS size. A novel nonlinear method for BESS optimal sizing in a grid-independent MG was presented in [26]. In addition to the nonlinearity, DOD was also not considered. Recognizing various roles of different stress factors on battery degradation, [27] proposed an accelerated testing approach to model degradation. The nonlinearity and complexity of the model, however, significantly increases the computation time, and a planning model based on a nonlinear mathematical model may not converge with the optimal solution. Considering the battery degradation cost, [28] discussed optimal BESS sizing. However, it does not model PV asset degradation in the long-term planning model.

### *1.2 Contributions*

As discussed, the existing body of literature highlights the following key gaps in MG planning models:

1. Omission of BESS and PV degradation: Many studies neglect the impact of degradation on determining the optimal sizes of energy resources, leading to unrealistic long-term planning outcomes.
2. Challenges with nonlinear models: Nonlinear approaches often suffer from long solving times and convergence to locally optimal solutions. Simplified models that use reduced temporal resolutions to address these issues compromise solution reliability.
3. Static degradation cost modeling for BESS: Traditional models use fixed penalty costs for BESS degradation, failing to account for DOD variations, resulting in inaccurate cost estimations and under-utilization of BESS assets.

This paper addresses these gaps through the following key contributions:

1. Iterative Post-Optimization Correction (IPOC) Framework: A novel approach that dynamically aligns BESS degradation costs with observed DOD profiles, iteratively refining cost calculations to improve both accuracy and asset utilization.
2. Integration of PV degradation costs: The model incorporates PV degradation into long-term MG planning, ensuring a realistic and comprehensive evaluation of asset performance.
3. MILP-based approach for scalability: By utilizing a mixed-integer linear programming (MILP) formulation, the model achieves computational efficiency, scalability, and global optimality.
4. Capability for high-resolution data handling: The model is capable of processing high-resolution temporal data without requiring reduced

resolutions or non-linear complexity, ensuring reliability and accuracy.

Results showcase substantial cost savings, enhanced PV utilization, and overall model accuracy, positioning this methodology as a robust, scalable solution for sustainable MG planning. Section II of this paper presents the optimization models for MG resource sizing problems. Section III presents the proposed methodology with section IV providing the case studies. Finally, section V draws the conclusions.

## 2. Optimization Problem Formulation

For typical MG planning projects, the primary objective is to minimize total costs, encompassing: (1) initial capital investment, (2) maintenance and operational expenses, and (3) degradation/replacement costs, if applicable. The supplementary aim of this study is to highlight the added benefits of incorporating a BESS, even with its associated degradation and replacement costs. To achieve this, three MG resource sizing models are implemented. The primary benchmark case considers a controllable DG source and a PV system as the only candidate resources, excluding BESS. This model optimizes the sizes of PV and DG sources to maintain the power balance in the planning period and is referred to as the microgrid sizing (MGS) model. An ideal BESS is then added to the benchmark MGS model, referred to as MGS-IB, which serves as a secondary benchmark by neglecting any degradation costs associated with the BESS. Finally, the MGS model with a non-ideal BESS (MGS-NIB) is developed to capture the practical scenario of a BESS undergoing degradation due to operational use. All three models include PV asset degradation, which is primarily influenced by environmental factors such as ultraviolet radiation, thermal cycling, and humidity [29]. Unlike BESS, PV degradation is not linked to operation and is typically represented as an annual rate. Its degradation cost is modeled as a fraction of the initial investment, depending on system scale.

In this study, three distinct simulation resolutions are used to ensure comprehensive evaluation: (i) 365-day-resolution: it covers every single day of the entire year, with each day having a different hourly load and solar power profile; (ii) 12-day-resolution: it encompasses 12 typical days representing a year, with each day encapsulating an averaged hourly load and solar power profile for each month; (iii) 1-day-resolution: it only uses a single representative day and assumes every day has the same hourly load and solar power profile throughout the planning duration.

### *2.1 Benchmark MGS Model*

The objective function for the benchmark MGS model is shown in (1). The MGS model's objective function ($f_1$) consists of the DG operational and no-load costs, DG capital cost, PV capital cost, and the PV degradation cost. Note that the capital costs are just one-time costs, but the operational and degradation costs are assumed to be repeated for each year in the MG planning period.

$$\begin{aligned} min\ f_1 = \alpha . Y_{MG} . \sum_{d \in D} \sum_{t \in T} (P_{d,t}^{DG} . C_{DG}^{op} + U_{d,t}^{DG} . C_{DG}^{nl}) + \\ P_{DG}^{max} . C_{DG}^{capital} + S_{PV} . C_{PV}^{capital} + C_{PV}^{deg} . Y_{MG} \end{aligned} \tag{1}$$

Constraints of the MGS model are shown below:

$$P_{d,t}^{DG} = P_{d,t}^{load} - (P_{d,t}^{PV} . S_{pv} - P_{d,t}^{pv_{curt}}) \tag{2}$$

$$P_{d,t}^{DG} \leq P_{DG}^{max} . U_{d,t}^{DG} \tag{3}$$

$$P_{d,t}^{DG} \geq P_{DG}^{min} . U_{d,t}^{DG} \tag{4}$$

$$P_{d,t}^{pv_{curt}} \leq P_{d,t}^{PV} . S_{PV} \tag{5}$$

$$C_{PV}^{deg} = \gamma_{PV}^{rep} . \left( C_{PV}^{capital} . S_{PV} . \delta_{PV}^{deg} \right) \tag{6}$$

Whether it is a bulk power system or an MG, maintaining the load and source power balance is crucial for operation. Constraint (2) enforces the power balance between generation and demand for the benchmark optimization model. The DG's maximum and minimum power limits are enforced by (3) and (4). Equation (5) models the solar power curtailed during periods of more solar generation than the load. The degradation cost of PV is calculated by (6). Note that the degradation cost is dependent on the size of the PV system and is a fraction of the original capital cost. As explained earlier, the degradation rate is constant per annum.

### *2.2 MGS-IB Model*

The objective function for the secondary benchmark MGS-IB ($f_2$) model is shown in (7). In addition to the cost terms in (1), MGS-IB's objective includes the capital cost of BESS.

$$min\ f_2 = f_1 + S_{BESS} . C_{BESS}^{capital} \tag{7}$$

The constraints of the MGS-IB model that are exclusive to it are described from (8)-(17). In contrast to the MGS model, the power balance constraint involves the BESS charging and discharging power and is expressed in (8). Constraints (9) and (10) restrict the BESS energy level to remain in the allowable range of state of charge (SOC). Equation (11) models the BESS's non-simultaneous charging and discharging behavior. Similar to the DG, constraints (12) and (13) restrict the BESS output power to stay within limits.

These constraints also govern the C-rate of the battery, where $T_{BESS}^{chg}$ and $T_{BESS}^{dchg}$ represent the time required to fully charge or discharge the BESS from 0% to 100% SOC or vice versa. Equations (14) calculates and updates the BESS energy levels for each time period based on its charging and discharging behavior. To ensure the BESS returns to its starting energy level at the end of the simulation resolution, constraints (15) and (16) are used. Equation (15) is applicable to non-365-day resolutions, such as 1-day and 12-day resolutions tested in this work, where days are treated as independent, and the SOC does not carry over between them. Conversely, equation (16) applies to the 365-day resolution, where days are continuous, and the SOC naturally flows from the end of one day to the start of the next.

$$P_{d,t}^{DG} = P_{d,t}^{load} + P_{d,t}^{chg} - P_{d,t}^{dchg} - (P_{d,t}^{PV}.S_{pv} - P_{d,t}^{pv_{curt}}) \tag{8}$$

$$SOC_{min}.S_{BESS} \leq E_{d,t}^{BESS} \leq SOC_{max}.S_{BESS} \tag{9}$$

$$SOC_{min}.S_{BESS} \leq E_{BESS}^{init} \leq SOC_{max}.S_{BESS} \tag{10}$$

$$U_{d,t}^{chg} + U_{d,t}^{dchg} \leq 1 \tag{11}$$

$$P_{d,t}^{chg} \leq U_{d,t}^{chg}.\frac{S_{BESS}}{T_{BESS}^{chg}} \tag{12}$$

$$P_{d,t}^{dchg} \leq U_{d,t}^{dchg}.\frac{S_{BESS}}{T_{BESS}^{dchg}} \tag{13}$$

$$E_{d,t}^{BESS} = E_{d,t-1}^{BESS} + \left(\eta_{BESS}.P_{d,t}^{chg} - P_{d,t}^{dchg}\right), 1 \leq t \leq N_T \tag{14}$$

$$E_{d,N_T}^{BESS} = E_{BESS}^{init}, \forall d \in D \tag{15}$$

$$E_{N_D,N_T}^{BESS} = E_{BESS}^{init} \tag{16}$$

### *2.3 MGS-NIB Model*

The objective function for this model ($f_3$) is shown in (17). In addition to all the terms in MGS-IB model, MGS-NIB's objective function aggregates the degradation cost of BESS.

$$min\ f_3 = f_2 + C_{BESS}^{deg}.\alpha.Y_{MG} \tag{17}$$

The constraints exclusive to this model are presented below:

$$\delta_{BESS}^{deg} = \frac{C_{BESS}^{capital}.S_{BESS}.\gamma_{BESS}^{rep}}{S_{BESS}.CL_{cycles}^{BESS}} \tag{18}$$

$$C_{BESS}^{deg} = \delta_{BESS}^{deg}.\sum_{d \in D}\sum_{t \in T} P_{d,t}^{dchg} \tag{19}$$

Constraint (18) calculates the degradation cost factor of the BESS with (19) computing the total degradation cost based on the discharge energy throughput, which is grounded on the heuristic linear battery degradation model [30]-[31]. The numerator in (18) indicates the total amount in dollars to replace the BESS of a certain MWh capacity. The denominator represents the total available energy in MWh based on the cycle life at a certain DOD level. Equation (18) automatically reduces to (20), thereby eliminating the inherent nonlinearity. The resultant relation implies that the cost of battery degradation is strongly dependent on the cycle life. Also, note that BESS degradation is modeled as a penalty factor in the objective function. Thus, a higher penalty would restrict the battery usage to keep the cost of degradation minimal.

$$\delta_{BESS}^{deg} = \frac{C_{BESS}^{capital} \cdot \gamma_{BESS}^{rep}}{CL_{cycles}^{BESS}} \tag{20}$$

The nonlinearities in (3), (4), (12), and (13) are eliminated using the BigM method [32] and the resultant equations are presented below:

$$P_{d,t}^{DG} \leq BigM. U_{d,t}^{DG} \tag{21}$$

$$P_{d,t}^{DG} \leq P_{DG}^{max} \tag{22}$$

$$P_{d,t}^{DG} \geq BigM. U_{d,t}^{DG} \tag{23}$$

$$P_{d,t}^{DG} \geq P_{DG}^{max} \tag{24}$$

$$P_{d,t}^{chg} \leq BigM. U_{d,t}^{chg} \tag{25}$$

$$P_{d,t}^{chg} \leq \frac{S_{BESS}}{T_{BESS}^{chg}} \tag{26}$$

$$P_{d,t}^{dchg} \leq BigM. U_{d,t}^{dchg} \tag{27}$$

$$P_{d,t}^{dchg} \leq \frac{S_{BESS}}{T_{BESS}^{dchg}} \tag{28}$$

Table 1 presents the optimization model-to-equations mapping table that summarizes the respective constraints used to run the respective model.

Table 1. Optimization Equation-Model Mapping Table

| *No.* | *Model* | *Utilized Equations* |
|---|---|---|
| 1 | MGS | (1), (2), (5), (6), & (21)-(24) |
| 2 | MGS-IB | (5)-(11), (14)-(16), & (21)-(28) |
| 3 | MGS-NIB | (5), (6), (8)-(11), (14)-(17), & (19)-(28) |

## 3. Iterative Post Optimization Correction

The modeling of degradation cost of the BESS, as depicted by (19) and (20), is modeled assuming a fixed degradation cost factor. This factor serves as a penalty term for BESS operation. Intrinsically, this penalty factor is tied to the cycle life of the battery, which is heavily influenced by the DOD level. It's important to highlight that in real-world scenarios, a battery may not consistently maintain a constant DOD unless it is deliberately enforced. This observation underscores the need for an evaluation of the SOC profile of the battery. Such an evaluation can reveal the total number of cycles at each DOD level, along with their respective degradation costs. Fig. 1 offers a visual representation of this concept, illustrating the operation of a battery over a single day. The SOC profile in this figure reveals approximately three distinct discharging cycles occurring at different times throughout the day. Each of these cycles is characterized by a unique DOD magnitude. The variability in DOD levels across different cycles can impact the overall degradation cost and the optimization of the BESS operation. Therefore, understanding these dynamics is key for accurately predicting the BESS performance and informing efficient energy storage management strategies.

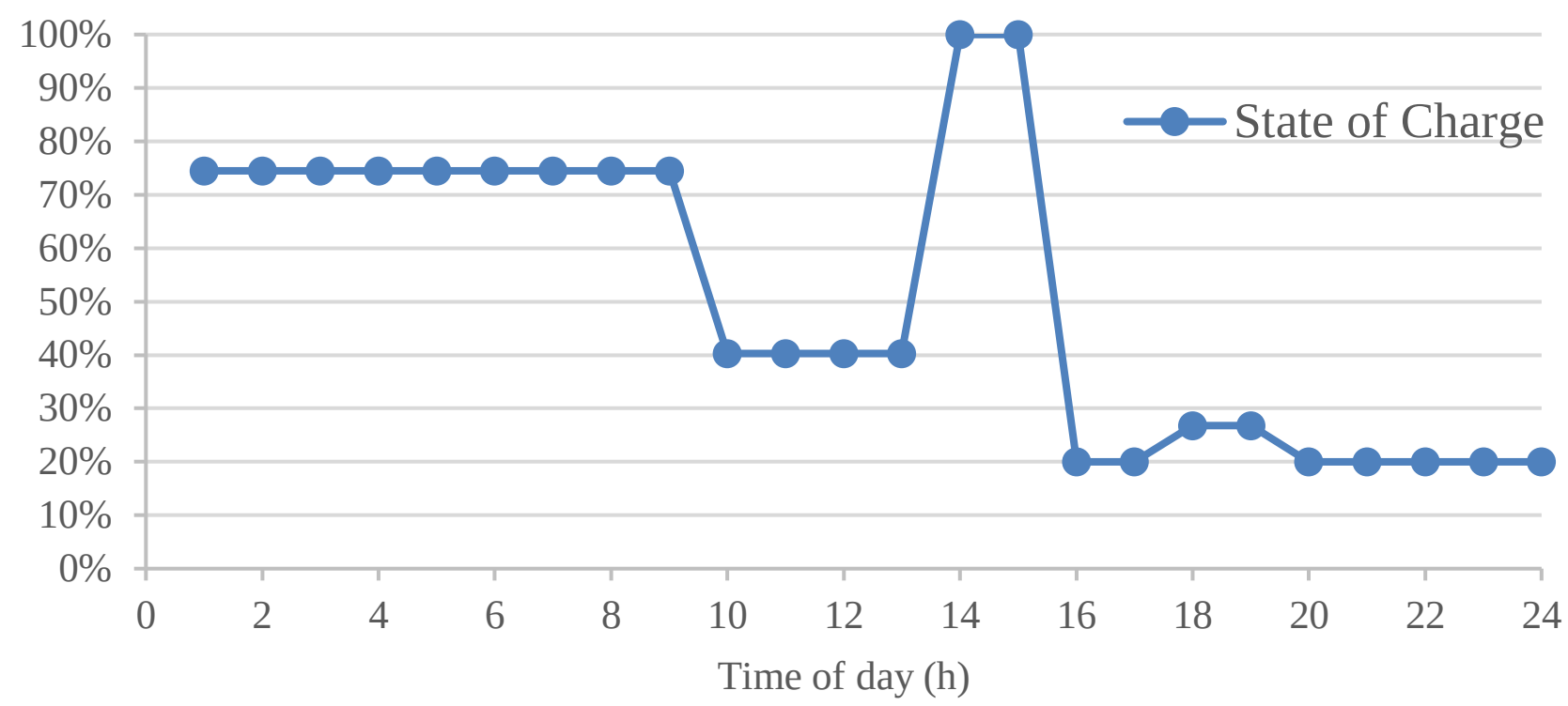

Fig. 1. BESS SOC profile illustrating discharging cycles at different DOD levels

The results obtained with a fixed penalty cost can thus over/under utilize the BESS potential in MG planning and resource scheduling problem. To improve the accuracy of the results, we propose an IPOC algorithm for refining the BESS degradation cost. The actual degradation cost of the BESS can be approximated by (29) and the correction involves subtracting this value from equation (19). This iterative methodology contributes to the precision and reliability of the overall modeling outcomes.

$$C_{BESS}^{deg_{actual}} = \sum_{i=1}^{n} count_{DOD_i}.\delta_{BESS_i}^{deg}.DOD_i.S_{BESS} \quad (29)$$

Here 'n' represents the total number of DOD levels, and 'i' denotes the index of the DOD level. This formulation integrates the degradation cost per cycles count across all DOD levels. For subsequent iterations in the optimization model, the denominator in equation (20) is updated based on the actual battery usage profile, using the average DOD within the given time set as the iterative variable. The algorithm's methodology is briefly depicted in the flowchart presented in Fig. 2.

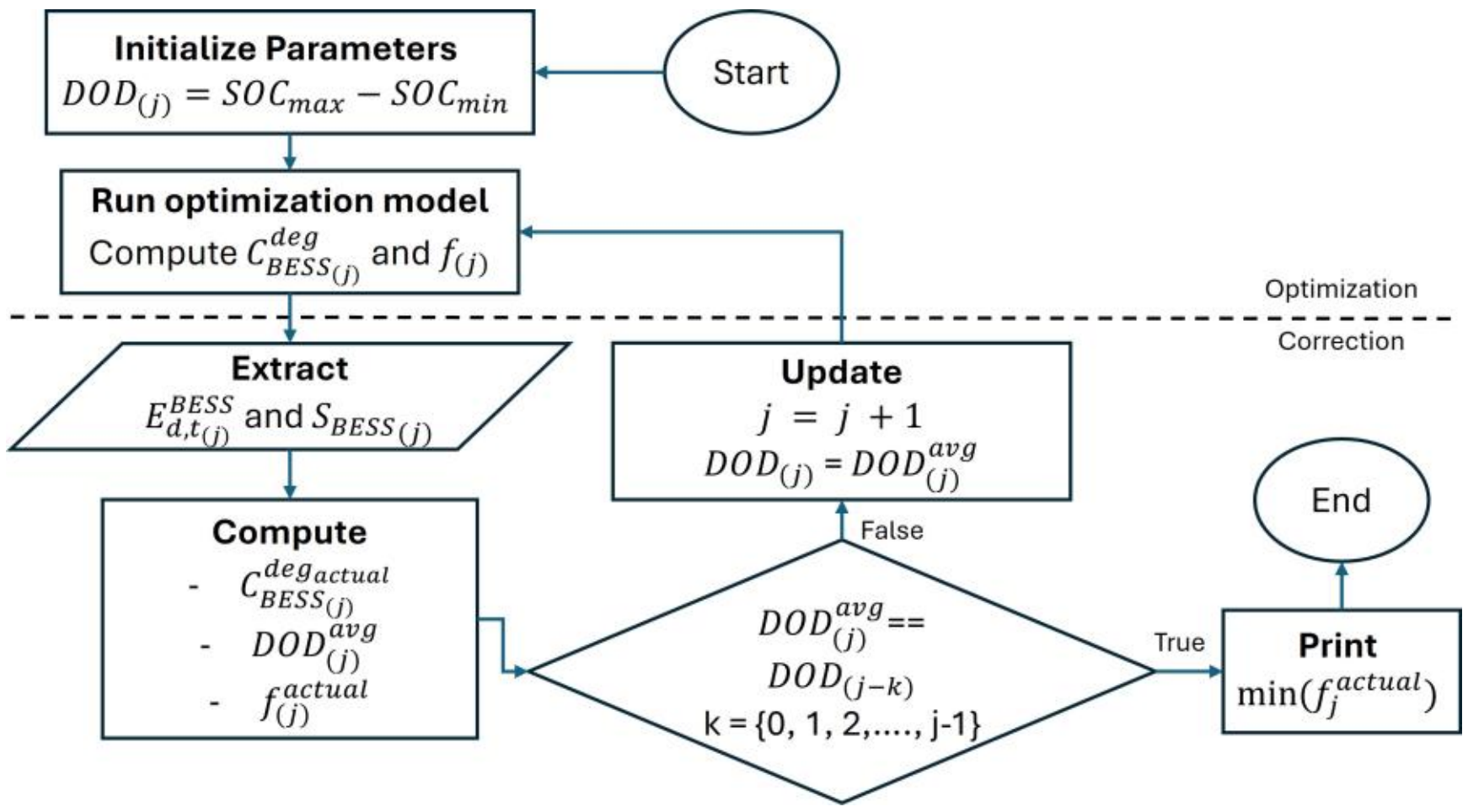

Fig. 2. Proposed methodology framework for IPOC

The process begins with the initialization of parameters, where the $DOD_{(j)}$ represents the depth of discharge for the iteration 'j'. and is calculated by taking the difference between the maximum and minimum allowable SOC levels of the BESS. This value influences the degradation cost factor, which serves as a penalty term in the optimization function. During each iteration 'j', the optimization model is solved to compute two outputs, the BESS degradation cost $C^{deg}_{BESS_{(j)}}$ and the total objective cost $f_{(j)}$. Since the BESS does not always cycle to a specific DOD level, the initially calculated degradation cost might not be accurate. The proposed IPOC methodology addresses this by first extracting the operational profile of the BESS and calculates SOC by dividing the time-series energy data $E^{\text{BESS}}_{d,t_{(j)}}$ with the BESS size $S_{BESS_{(j)}}$. This data is then processed into a separate function to compute the actual BESS degradation cost $C^{deg_{actual}}_{BESS_{(j)}}$ using the equation (29), along with actual average DOD denoted as $DOD^{avg}_{(j)}$, as well as the actual objective cost $f^{actual}_{(j)}$. If the actual average DOD matches the assumed DOD or aligns with a DOD value from any previous iteration ($DOD_{j-k}, k = 0, 1, \dots, j-1$), the algorithm terminates and outputs the minimum objective cost solution min ($f^{actual}_{(j)}$) as the global optimum solution. Otherwise, the iteration index is incremented, and the observed average DOD is set as the new assumed DOD for the next iteration and is fed back into the optimization model. The iterative nature of the algorithm ensures that the degradation cost dynamically adapts to the operational behavior of the BESS, improving the accuracy of cost estimations while avoiding under-utilization and maintaining computational efficiency. The framework effectively links operational metrics such as SOC, DOD, and degradation costs to provide a robust, scalable, and reliable methodology for optimizing BESS integration.

## 4. Case Studies

A test system opted for this study is illustrated in Fig. 3. It describes a remote MG that is planned to be installed in a certain location to serve a residential community. The test system comprises a solar PV farm and an open-cycle NG power plant acting as a DG source and serves as a backup and load-balancing resource when PV output or BESS capacity is insufficient. The energy storage system elected is a lithium iron phosphate (LiPO4) based BESS which is normally the standard choice for energy intensive applications. The load data is taken from the OpenEI TMY2 commercially available residential load database for Houston, Texas (29.7°N, 95.4°W) and is downscaled to a load profile having a peak load of 0.8 MW, 0.05 MW as the minimum load, and an average load of 0.17 MW [33]. In addition, the MG planning horizon is taken to be 25 years that represents an average power plant lifetime period.

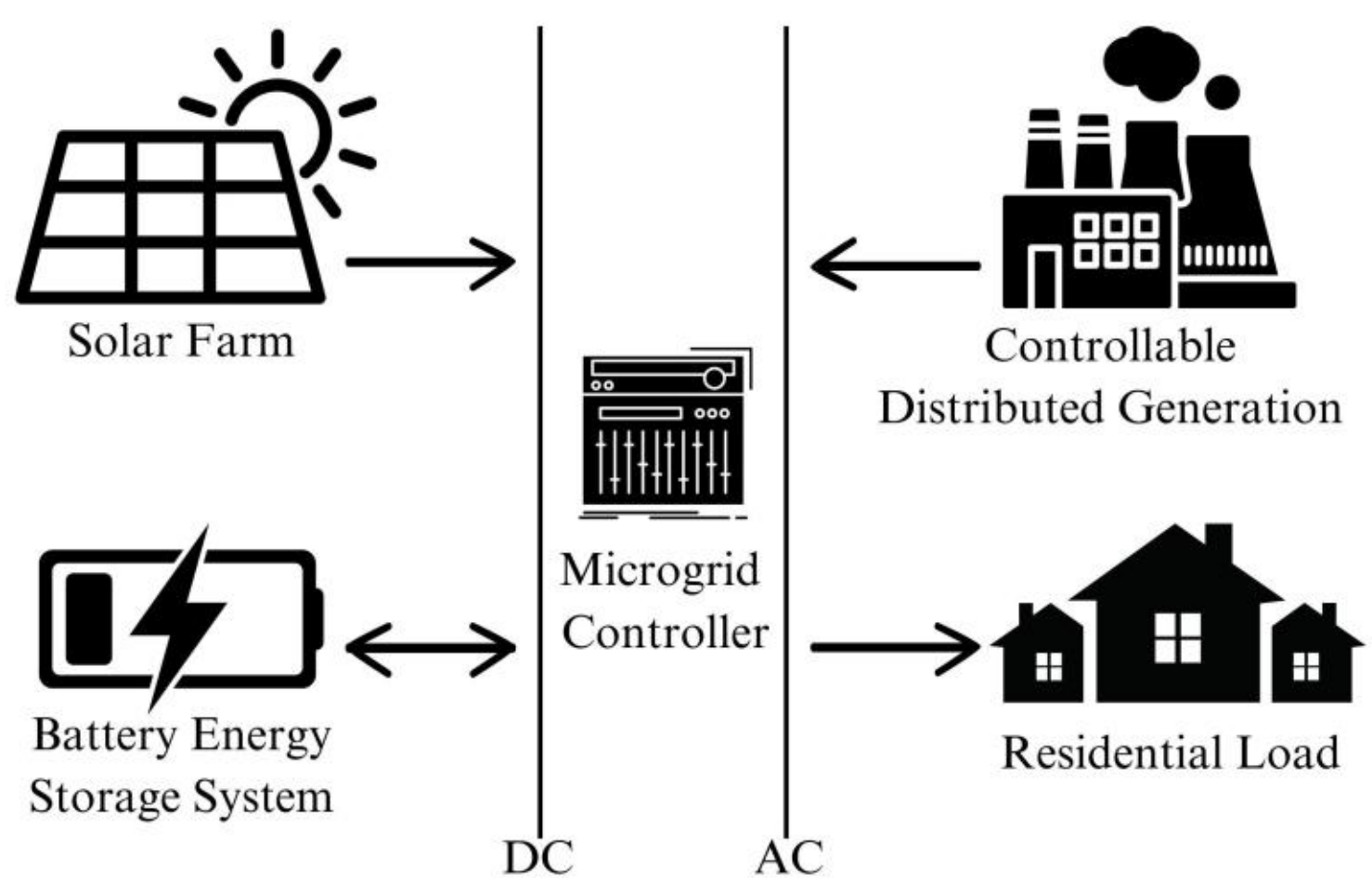

Fig. 3. Test system overview

Fig. 4 represents the daily load profile at a one-hour resolution, alongside the solar power production profile of a standard 1 MW PV system. These profiles are derived from a comprehensive dataset that spans yearlong hourly data which are averaged for the same hours of load and PV for the said latitude and longitude. The PV data is obtained using the National Renewable Energy Lab's PVWatts tool [34], configured as per the stated geographical coordinates. This ensures that the PV data accurately reflects the solar energy potential of the specific location, considering local weather patterns and solar irradiance levels.

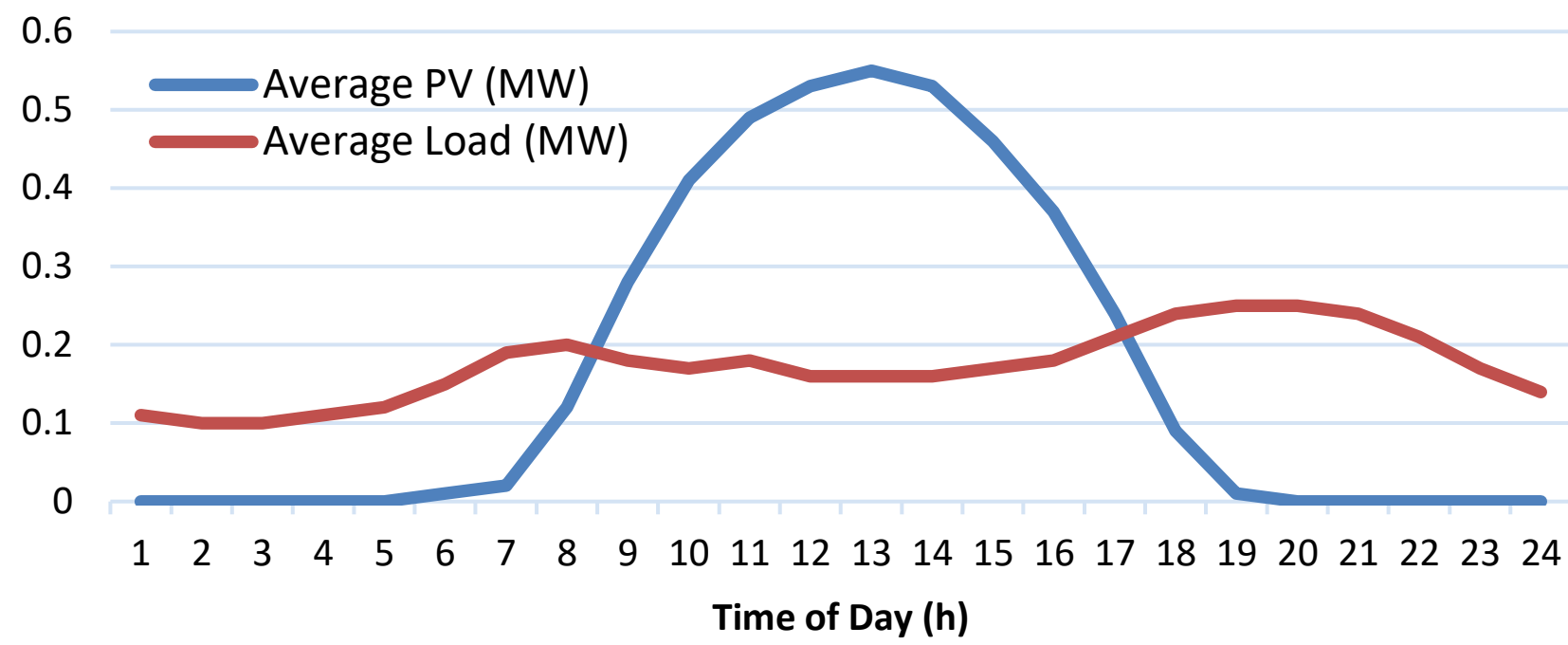

Fig. 4. Average daily load and PV power profiles.

The characteristics of the selected PV farm and NG generator are shown in tables 2 and 3 respectively that are extracted from [35]-[39]. Drawing from publicly accessible data sources is an indicative of the research's practicality.

Table 2. PV Characteristics

| *Capital cost* | *Replacement cost* | *PV degradation rate* |
|---|---|---|
| 1450,000 $/MW | 41% of capital | 1% per annum |

Table 3. DG Characteristics

| *Capital cost* | *Operational cost* | *No load cost* |
|---|---|---|
| 1150,000 $/MW | 44.75 $/MWh | 5.25$/h |

Table 4 illustrates the characteristics of the selected BESS extracted from [37],[40]-[41]. Battery degradation is subject to fluctuations in charge/discharge rates, operating temperatures, DOD levels, and the operating SOC range. Cell manufacturers market their cells in diverse operational configurations to accommodate these variations. The specifics of the BESS parameters are crucial considerations in a study aiming to perform its optimal sizing.

Table 4. BESS Characteristics

| *Roundtrip η* | *C-rate* | *Capital cost* | *Replacement cost* |
|---|---|---|---|
| 90% | 1C | 469,000$/MWh | 79% of capital |

Within the scope of this paper, it is assumed that the BESS operating temperature is controlled by a thermal management system maintaining it at around room temperature (25°C). Table 5 shows the relationship between DOD and cycle life for a commercially available battery pack [42]. It is important to acknowledge the practical challenges faced by cell manufacturers in disseminating cycles to failure versus DOD data across multiple DOD levels as the process is considerably energy and time intensive. Consequently, this study employs interpolation techniques on the acquired data to overcome these practical limitations and serves as a fair approximation.

Table 5. BESS DOD vs Cycle life data

| *DOD %* | *10-20* | *30-40* | *50-60* | *70-80* | *90-100* |
|---|---|---|---|---|---|
| *Cycles* | 14500 - 12000 | 9600 - 7500 | 5800 - 4600 | 3400 - 3000 | 2200 – 2000 |

The three MG sizing models are executed on the same test system. The selection of simulation resolution is handled by the scaling factor ‘α’ that denotes the repetition of load and solar power profiles in a year. Thus, for the 365-day-resoultion, it is set to 1, for the 12-day-resoultion, its value is 30.42, which is the average number of days per month in a year. Finally, for the 1-day-resolution, α takes on the value of 365. Obviously, the 365-day-resolution case has the full resolution, and the associated results would be most convincing, while the 1-day-resolution case would be least accurate, but it may require the least computing resources to solve.

The simulations are conducted using AMPL software, executed on an Intel(R) Xeon(R) W-2195 CPU 2.30GHz processor with 128GB RAM. The GUROBI solver is employed with a MIPGAP setting of 0.0 to ensure optimal solutions and a TIME LIMIT setting of 43,200 seconds to handle computationally intensive cases.

Table 6 captures a comprehensive cost comparison over the three optimization models, using the most accurate 365-day-resolution. Additionally, Table 7 details the comparative analysis of plant sizes, while Table 8 presents the share of energy utilization.

Table 6. All costs comparison (365-day-resolution)

| *Attribute* | *MGS* | *MGS-IB* | *MGS-NIB* |
|---|---|---|---|
| Solution | Optimal | Feasible | Feasible |
| Computational time (s) | 3.4 | 43,290 | 43,739 |
| Relative optimality gap | ~0% | ~3.5% | ~2% |
| Objective cost ($M) | 3.663 | 2.982 | 3.551 |
| DG capital cost ($M) | 0.920 | 0.689 | 0.713 |
| PV capital cost ($M) | 0.483 | 0.453 | 0.463 |
| BESS capital cost ($M) | - | 0.211 | 0.174 |
| DG OP+NL cost ($M) | 2.210 | 1.583 | 2.026 |
| PV degradation cost ($M) | 0.050 | 0.046 | 0.047 |
| BESS degradation cost ($M) | - | - | 0.128 |

Table 7. Optimal plant size comparison (365-day-resolution)

| *Attribute* | *MGS* | *MGS-IB* | *MGS-NIB* |
|---|---|---|---|
| DG size (MW) | 0.80 | 0.60 | 0.62 |
| PV size (MW) | 0.33 | 0.31 | 0.32 |
| BESS size (MWh) | - | 0.45 | 0.37 |

Table 8. Energy utilization comparison (365-day-resolution)

| *Attribute* | *MGS* | *MGS-IB* | *MGS-NIB* |
|---|---|---|---|
| Total load (MWh) | 37,863.8 | 37,863.8 | 37,863.8 |
| DG generation (MWh) | 27,737.0 | 28,259.5 | 26,798.8 |
| PV generation (MWh) | 12,435.5 | 11,643.4 | 11,912.8 |
| PV curtailed (MWh) | 2,308.75 | 371.8 | 709.3 |
| BESS discharging (MWh) | - | 15,006.8 | 1,247.2 |

The results reveal a noteworthy trend in the objective costs, as depicted in Table 6 and Figure 5. The absence of a BESS in the microgrid setup, represented by the benchmark MGS model, corresponds to the highest objective cost, at $3.663 million (M). However, the introduction of an ideal BESS in the MGS-IB model leads to a significant ~18.6% reduction in cost, highlighting the potential cost-saving benefits of battery integration. The MGS-NIB model, which incorporates a more realistic BESS configuration accounting for degradation, yields a more pragmatic ~3.1% cost reduction. These trends are clearly illustrated in Figure 5, which compares the objective costs of all three models, emphasizing the trade-offs between ideal and non-ideal BESS implementations. Table 7 and Figure 6 present the optimal plant size comparisons across the models. In the MGS model, the DG capacity is 0.80 MW, while PV capacity is 0.33 MW. The MGS-IB model, with an ideal BESS, optimizes these values further, reducing DG size to 0.60 MW and PV size to 0.31 MW while adding 0.45 MWh BESS. In contrast, the MGS-NIB model balances the inclusion of a non-ideal BESS (0.37 MWh) by slightly increasing the DG size to 0.62 MW and PV size to 0.32 MW. These shifts highlight the system's response to different BESS configurations and their impact on plant sizing. Table 8 and Figure 7 provide an energy utilization comparison. A key finding is the remarkable reduction in solar energy curtailment when a BESS is integrated. The MGS model curtails ~2,309 MWh of solar energy, whereas the MGS-IB model reduces this by ~84% to just 372 MWh, and the MGS-NIB model achieves a ~69% reduction, curtailing 709 MWh in the planning period. These reductions underscore the importance of BESS in enhancing solar utilization and reducing reliance on DG generation. Additionally, the energy discharged from the BESS in the MGS-IB and MGS-NIB models further emphasizes the role of BESS in load management.

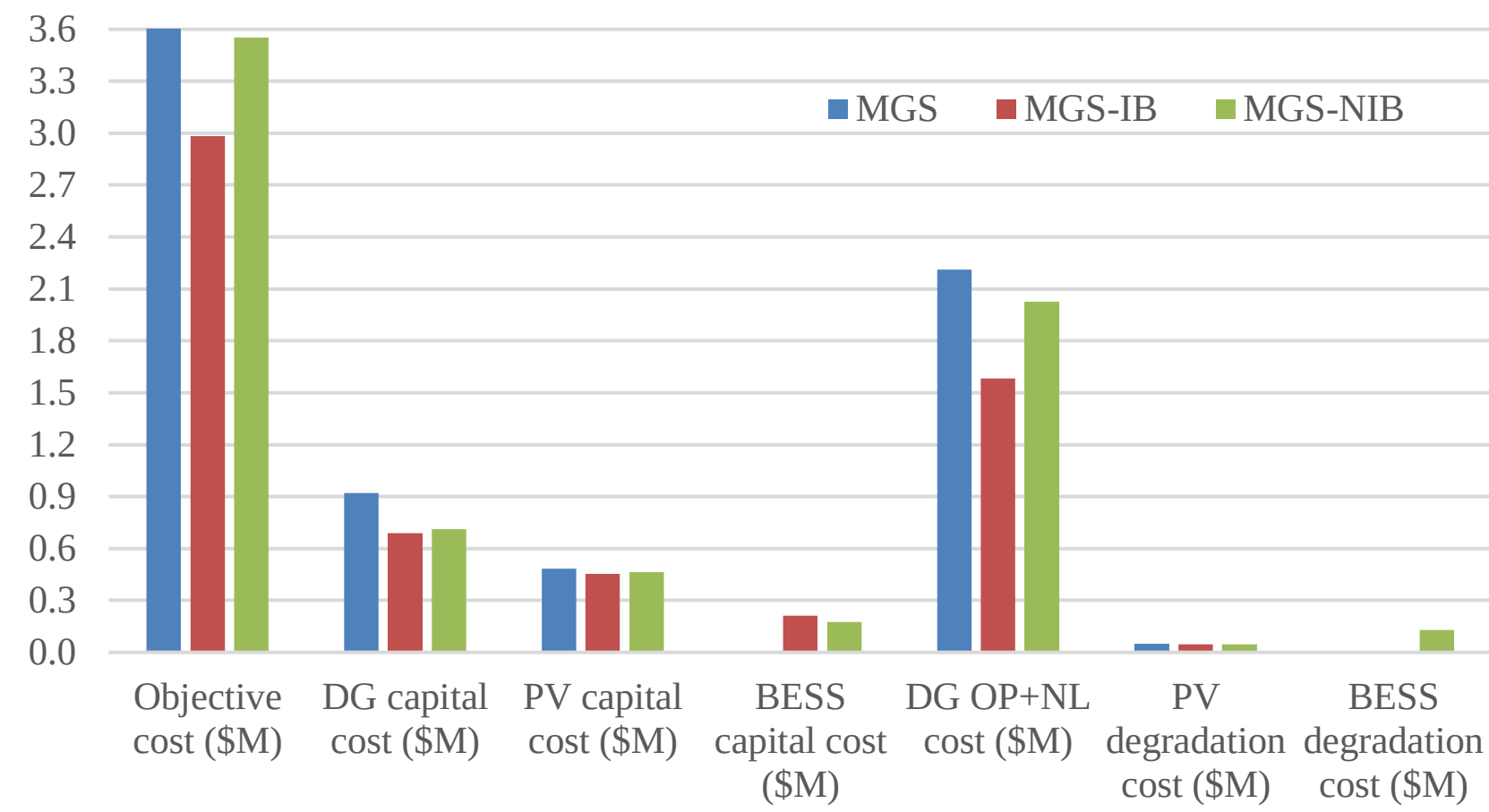

Fig. 5. Comparison of costs (365-day-resolution)

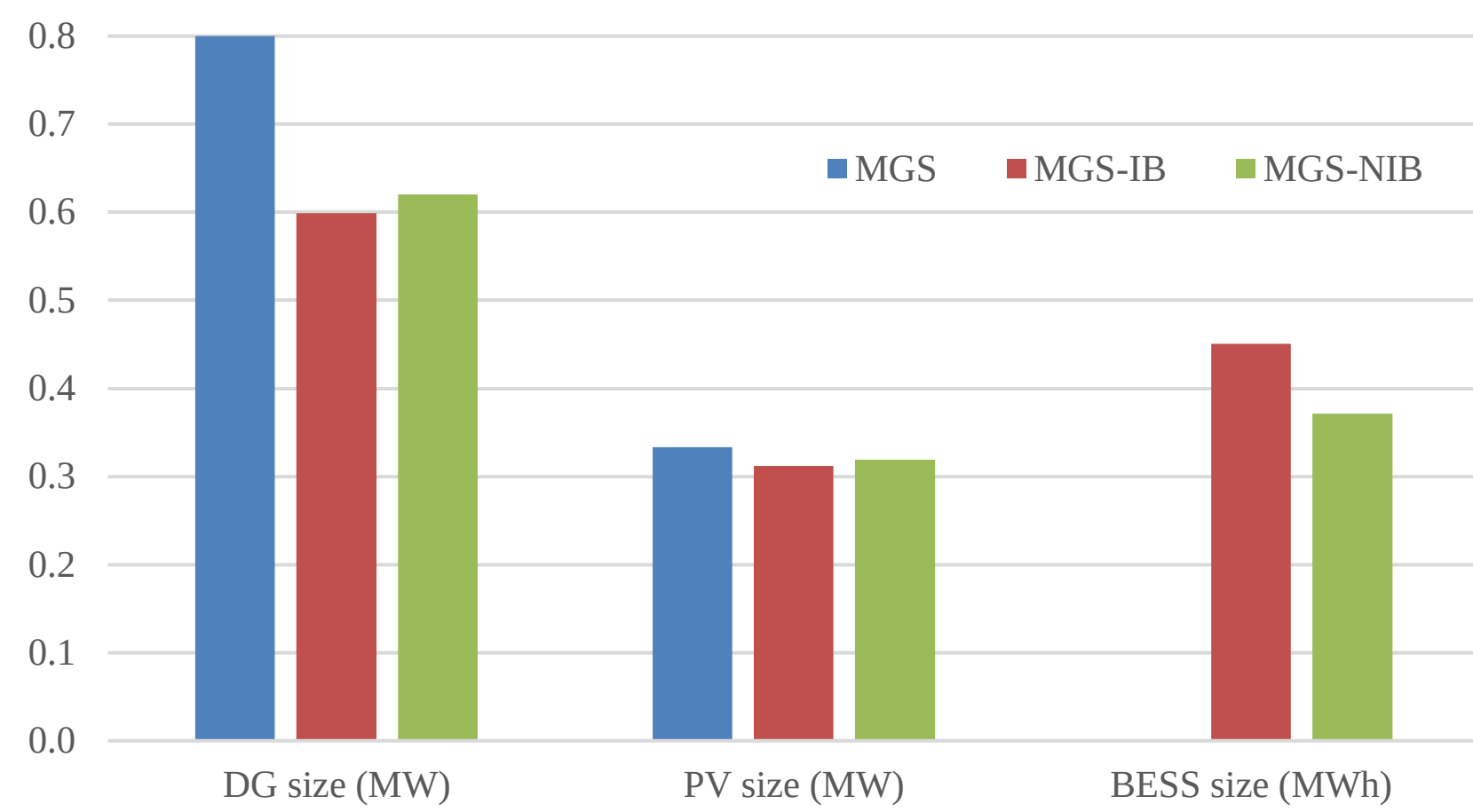

Fig. 6. Comparison of optimal plant sizes (365-day-resolution)

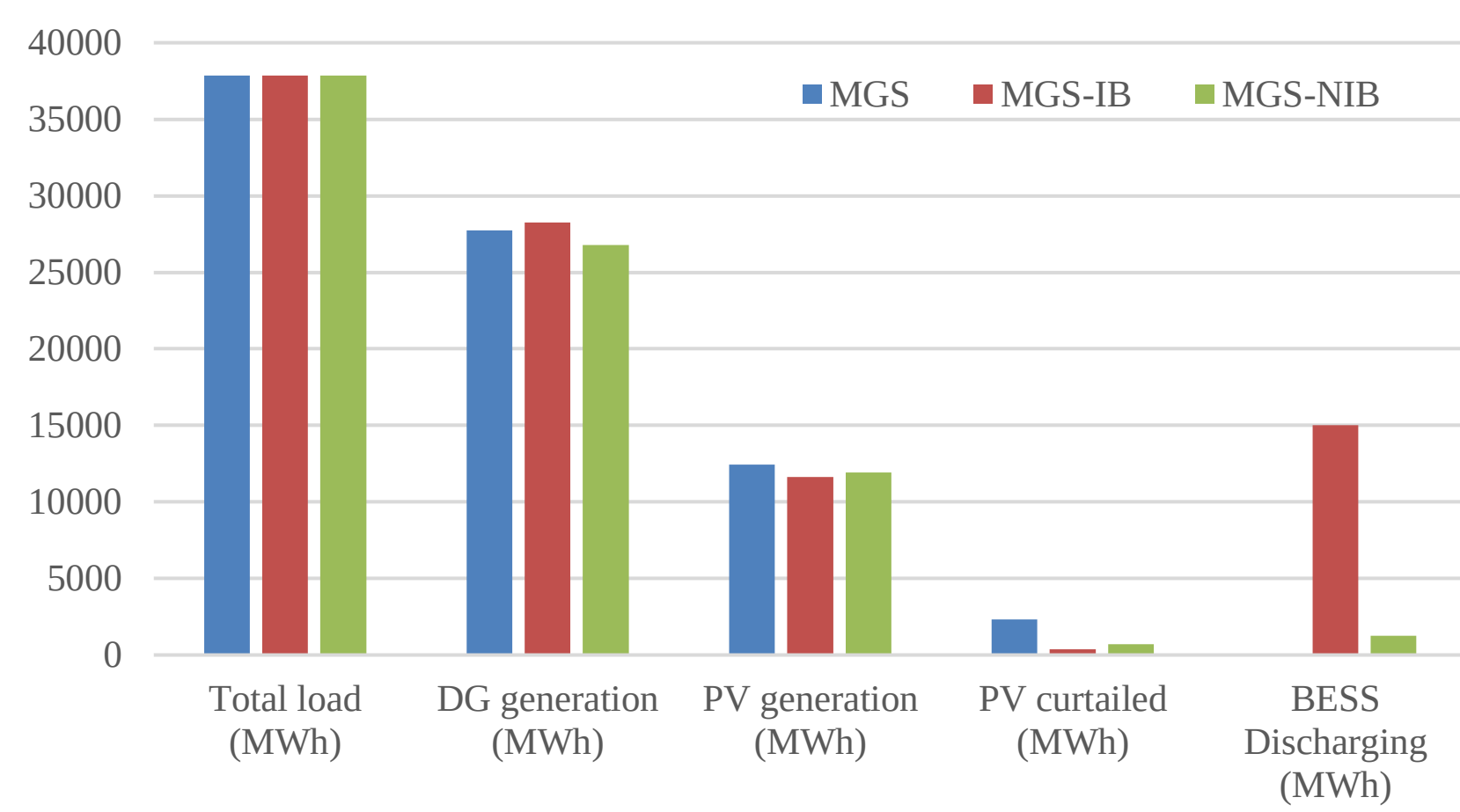

Fig. 7. Comparison of energy utilizations (365-day-resolution)

Table 9 provides a similar assessment of all costs embedded in the model, employing the 12-day resolution for a detailed analysis. Furthermore, the comparison of plant sizes is presented in Table 10, while Table 11 illustrates the share of energy utilization, both within the same temporal framework.

Table 9. All costs comparison (12-day-resolution)

| *Attribute* | *MGS* | *MGS-IB* | *MGS-NIB* |
|---|---|---|---|
| Solution | Optimal | Feasible | Optimal |
| Computational time (s) | 0.203 | 43235.5 | 6.540 |
| Relative optimality gap | ~0% | ~3.9% | ~0% |
| Objective cost ($M) | 3.06 | 2.70 | 3.00 |
| DG capital cost ($M) | 0.37 | 0.31 | 0.34 |
| PV capital cost ($M) | 0.59 | 0.56 | 0.56 |
| BESS capital cost ($M) | - | 0.15 | 0.03 |
| DG OP+NL cost ($M) | 2.04 | 1.62 | 1.96 |
| PV degradation cost ($M) | 0.06 | 0.06 | 0.06 |
| BESS degradation cost ($M) | - | - | 0.05 |

Table 10. Optimal plant size comparison (12-day-resolution)

| *Attribute* | *MGS* | *MGS-IB* | *MGS-NIB* |
|---|---|---|---|
| DG size (MW) | 0.32 | 0.27 | 0.29 |
| PV size (MW) | 0.41 | 0.39 | 0.39 |
| BESS size (MWh) | - | 0.33 | 0.06 |

Table 11. Energy utilization comparison (12-day-resolution)

| *Attribute* | *MGS* | *MGS-IB* | *MGS-NIB* |
|---|---|---|---|
| Total load (MWh) | 379,77.8 | 379,77.8 | 379,77.8 |
| DG generation (MWh) | 252,69.9 | 247,06.6 | 252,80.9 |
| PV generation (MWh) | 152,09.2 | 144,55.1 | 144,78.8 |
| PV curtailed (MWh) | 2,501.27 | 465.1 | 1,723.8 |
| BESS discharging (MWh) | - | 6,469.0 | 523.5 |

These comprehensive tabulations extend our understanding of the model's performance under diverse time sets. The simulation results in this time set show a different percentage decrease in objective costs and solar energy curtailment. The addition of an ideal BESS indicates a decrease of ~12% in objective cost and ~81% in solar energy curtailment with respect to the MGS model. For the MGS-NIB model, the reduction in objective cost is ~2% and the solar energy curtailed is ~31%. These results suggest that using averaged profiles to represent an entire month of solar and load data is a reasonable approximation if computation time is of importance. However, for planning projects where rapid computations are not of key importance, higher resolution offers more accurate results.

Table 12 provides a comparable overview of all costs within the model, employing the 1-day-resolution. Additionally, Table 13 displays a comparison of plant sizes, while Table 14 outlines the proportional share of energy utilization, all based on the same time set.

Table 12. All costs comparison (1-day-resolution)

| *Attribute* | *MGS* | *MGS-IB* | *MGS-NIB* |
|---|---|---|---|
| Solution | Optimal | Optimal | Optimal |
| Computational time (s) | 0.03 | 0.45 | 0.13 |
| Relative optimality gap | ~0% | ~0% | ~0% |
| Objective cost ($M) | 2.906 | 2.595 | 2.880 |
| DG capital cost ($M) | 0.288 | 0.291 | 0.283 |
| PV capital cost ($M) | 0.601 | 0.604 | 0.536 |
| BESS capital cost ($M) | - | 0.125 | 0.025 |
| DG OP+NL cost ($M) | 1.956 | 1.513 | 1.920 |
| PV degradation cost ($M) | 0.062 | 0.062 | 0.055 |
| BESS degradation cost ($M) | - | - | 0.061 |

Table 13. Optimal plant size comparison (1-day-resolution)

| *Attribute* | *MGS* | *MGS-IB* | *MGS-NIB* |
|---|---|---|---|
| DG size (MW) | 0.25 | 0.25 | 0.25 |
| PV size (MW) | 0.41 | 0.42 | 0.37 |
| BESS size (MWh) | - | 0.27 | 0.05 |

Table 14. Energy utilization comparison (1-day-resolution)

| *Attribute* | *MGS* | *MGS-IB* | *MGS-NIB* |
|---|---|---|---|
| Total load (MWh) | 37868.8 | 37868.8 | 37868.8 |
| DG generation (MWh) | 24430.5 | 23108.4 | 24695.7 |
| PV generation (MWh) | 15550.3 | 15621.5 | 13860.1 |
| PV curtailed (MWh) | 2112.10 | 0.0 | 620.7 |
| BESS discharging (MWh) | - | 7749.9 | 597.1 |

Results with this data resolution exhibit a consistent trend in objective cost and solar energy curtailment. However, it's crucial to note that these results, while following a similar pattern, are characterized by the lowest accuracy and the percentage savings in cost and may not be compelling enough for MG project investors.

Note that the simulation results assume a constant BESS degradation cost factor which is the penalty for cycling to 80% DOD. In actual, the battery may not be cycling to that level and thus requires post optimization data processing for correction. Table 15, 16, and 17 showcase the outcomes of the IPOC algorithm for the 365 days, 12 days, and 1-day data resolutions, respectively.

These tables encapsulate the refined results following the iterative correction process, providing a useful insight across different time resolutions.

Table 15. IPOC Results (365-day-resolution)

| *Attribute* | *Iteration 1* | *Iteration 2* |
|---|---|---|
| Average DOD: | 40% | 40% |
| Actual degradation cost ($M): | 0.091 | 0.254 |
| Cost correction ($M): | 0.037 | -0.090 |
| New objective cost ($M) | 3.514 | 3.531 |

Table 16. IPOC Results (12-day-resolution)

| *Attribute* | *Iteration 1* | *Iteration 2* |
|---|---|---|
| Average DOD: | 50% | 50% |
| Actual degradation cost ($M): | 0.037 | 0.145 |
| Cost correction ($M): | 0.017 | -0.028 |
| New objective cost ($M) | 2.983 | 2.974 |

Table 17. IPOC Results (1-day-resolution)

| *Attribute* | *Iteration 1* | *Iteration 2* | *Iteration 3* |
|---|---|---|---|
| Average DOD: | 40% | 70% | 40% |
| Actual degradation cost ($M): | 0.048 | 0.120 | 0.048 |
| Cost correction ($M): | 0.014 | -0.053 | 0.003 |
| New objective cost ($M) | 2.867 | 2.895 | 2.867 |

For the 365 days' time set, it can be seen that the objective cost of MGS-NIB model is ~4.1% less with respect to the objective cost of the benchmark MGS model, which in the pre-optimization model was yielding a reduction of ~3.1%. Similarly, for the 12 days' time set, the corrected objective cost is ~2.9% less with respect to the MGS model, which was originally 2.1% in the pre-optimization scenario. Lastly, the objective cost in the 1-day time set was originally ~1%, which got corrected to ~1.4% in the post-optimization model. These findings highlight the efficacy of the proposed IPOC algorithm in refining the planning model. The consistent improvements trend across varied time sets additionally indicate the algorithm's robustness in the enhancement of cost estimations.

Table 18 presents the sensitivity analysis results of the MGS-NIB model under varying economic assumptions, specifically focusing on ±20% changes in PV and BESS capital costs. This analysis aims to evaluate the robustness of the IPOC-based framework in adapting to different cost scenarios while maintaining optimal performance. Key metrics such as objective costs, iterations required for convergence, resource sizes, energy utilization, and observed average DoD are included to provide a comprehensive understanding of the model's adaptability.

Table 18. Sensitivity Analysis of MGS-NIB to PV and BESS Capital Costs (1-day-resolution)

| *Attribute* | *Benchmark Setting* | *PV Cost (-20%)* | *PV Cost (+20%)* | *BESS Cost (-20%)* | *BESS Cost (+20%)* |
|---|---|---|---|---|---|
| Original Objective cost ($M) | 2.880 | 2.754 | 2.998 | 2.863 | 2.889 |
| IPOC Objective Cost ($M) | 2.867 | 2.739 | 2.985 | 2.813 | 2.889 |
| Iterations | 3 | 2 | 3 | 4 | 1 |
| Avg. DoD | 40% | 50% | 40% | 40% | 80% |
| DG size (MW) | 0.246 | 0.25 | 0.25 | 0.24 | 0.25 |
| PV size (MW) | 0.370 | 0.41 | 0.37 | 0.37 | 0.37 |
| BESS size (MWh) | 0.054 | 0.03 | 0.05 | 0.14 | 0.02 |
| DG Gen (MWh) | 24695.7 | 24192.1 | 24695.7 | 24285.8 | 24899.3 |
| PV Gen (MWh) | 13860.1 | 15550.3 | 13860.1 | 13950.0 | 13860.1 |
| BESS usage (MWh) | 597.1 | 280.4 | 597.1 | 3303.3 | 168.6 |

When the PV capital cost decreases, leading to a corresponding reduction in its degradation cost, we observe a significant increase in PV resource allocation, accompanied by a decrease in the original objective cost. Conversely, when PV capital cost increases, the objective cost rises as expected due to the higher investment requirement. Similarly, for BESS, a decrease in its capital cost results in increased resource allocation and energy usage, followed by a reduction in the original objective cost compared to the benchmark. On the other hand, when BESS capital cost increases, resource allocation reduces, reflecting its reduced economic viability. In all these cases, except for the increased BESS cost scenario, the IPOC framework identified discrepancies between the observed DoD and the initially assumed DoD, necessitating revisions to the penalty terms. The application of IPOC revealed further reductions in objective costs, ranging from approximately 0.5% to 1.7% for the 1-day resolution. These improvements, which could be more pronounced under higher temporal resolutions, highlight the robustness of the IPOC framework. The results emphasize its effectiveness in refining objective costs by dynamically aligning the penalty term with actual operational profiles, ensuring accurate and reliable outcomes for diverse economic scenarios.

## 5. Conclusion

This paper presented a structured methodology for optimizing the integration of PV systems and BESS in MGs alongside a controllable DG source. The proposed MILP-based model effectively minimizes both capital and operational costs while addressing asset degradation costs. It ensures scalability, computational efficiency, and the ability to handle high-resolution temporal data. The proposed IPOC algorithm dynamically refines BESS degradation costs based on actual operational profiles. This approach overcomes the limitations of static degradation models, where overly low penalties can lead to overutilization and reliability concerns, while overly high penalties result in underutilization and suboptimal cost solutions. Additionally, by incorporating PV degradation costs, the model provides a realistic assessment of asset performance over long planning horizons.

For the 365-day time set, the MGS-NIB model without IPOC reduced the objective cost by ~3.1% compared to the benchmark MGS model. With the IPOC adjustments, the objective cost reduction increased to ~4.1%, showcasing a ~1% additional cost saving achieved through the iterative correction of degradation penalties. This trend is consistent across the 12-day and 1-day time resolutions, with varying levels of improvement. Sensitivity analysis further demonstrated the model's adaptability and robustness under diverse economic scenarios, with cost reductions ranging from 0.5% to 1.7% in the 1-day time set after IPOC adjustments. These findings underscore the economic and operational benefits of integrating BESS into microgrid systems, even when paired with low-operational-cost NG generators. While emissions constraints were not explicitly considered in this study, future research could explore their integration through approaches such as penalty costs or carbon credit markets. By addressing technical, economic, and operational considerations, the proposed methodology offers a robust and practical framework for sustainable microgrid planning, providing stakeholders with actionable insights to support the ongoing global transition to renewable energy.